\documentstyle[preprint,aps]{revtex}
\tightenlines

\newtheorem{theorem}{Theorem}
\newtheorem{acknowledgement}[theorem]{Acknowledgement}

\begin{document}
\title{Near Horizon Geometry of Extreme Black Holes and Colliding Waves}
\author{M. Halilsoy}
\address{Physics Department., Eastern Mediterranean University G.Magosa, North\\
Cyprus, Mersin 10, Turkey\\
email: mustafa.halilsoy@emu.edu.tr}
\date{\today }
\maketitle

\begin{abstract}
The correspondence between black holes and colliding waves extends to cover
the near horizon geometry of rotating black holes and colliding waves with
cross polarization. Extreme Kerr and Kerr-Newman geometries are given as
examples.
\end{abstract}

It is believed that matter collisions at speeds almost the speed of light
can produce mini black holes even in a laboratory. This possibility directs
attentions naturally on collisions of waves in general relativity. Local
equivalence (isometry) between two different types of spacetimes, such as
black holes (BHs) and colliding plane waves (CPWs) was pointed out first by
Chandrasekhar and Xanthopoulos CX [1]. This isometry covered only a portion
of the BH geometry, namely \ region between horizons and the interaction
region of CPWs. This concerns only those CPWs which create horizon instead
of an all encompassing space-like singularity. Such examples of CPWs are
very rare so far. The prototype is the solution of CX which describes
collision of impulse gravitational waves accompanied by shock gravitational
waves. As a second example we quote colliding electromagnetic (em) shock
waves of Bell and Szekeres (BS) [2], which is transformable to the
Bertotti-Robinson (BR) solution [3]. This latter is important from its
correspondence to conformal field theory and the underlying anti-de Sitter
(AdS) structure with a group $SL(2,R)\times SO(3).$ From geometrical point
of view BR represents the throat region of an extreme Reissner-Nordstrom BH
and its equivalence to colliding em waves is well-known [3].

In this letter we show that the connection between BHs and CPWs [4,5,6] is
not restricted only to the local isometry of CX, but seems to be more deep
rooted. This we do by showing that the near horizon geometry of the Extreme
Kerr and Kerr-Newman BHs are identical with particular CPWs.

The Kerr metric in the Boyer-Linquist coordinates is given by

\begin{equation}
ds^{2}=e^{2\nu }d\widetilde{t}^{2}-e^{2\psi }\left( d\widetilde{\phi }%
-\omega d\widetilde{t}\right) ^{2}-\rho ^{2}\left( \frac{d\widetilde{r}}{%
\widetilde{\Delta }}+d\theta ^{2}\right)
\end{equation}

where

\begin{eqnarray}
e^{2\nu } &=&\frac{\widetilde{\Delta }\rho ^{2}}{\left( \widetilde{r}%
^{2}+a^{2}\right) ^{2}-\widetilde{\Delta }a^{2}\sin ^{2}\theta } \\
e^{2\left( \psi +\nu \right) } &=&\widetilde{\Delta }\sin ^{2}\theta
\nonumber \\
\omega e^{-2\nu } &=&\frac{2Ma\widetilde{r}}{\widetilde{\Delta }\rho ^{2}}
\nonumber \\
\rho ^{2} &=&\widetilde{r}^{2}+a^{2}\cos ^{2}\theta  \nonumber \\
\widetilde{\Delta } &=&\left( \widetilde{r}-r_{+}\right) \left( \widetilde{r}%
-r_{-}\right)  \nonumber \\
r_{+} &=&M+\sqrt{M^{2}-a^{2}}  \nonumber \\
r_{-} &=&M-\sqrt{M^{2}-a^{2}}  \nonumber
\end{eqnarray}

In the extreme case we have $M=a,$ so that angular momentum is $J=M^{2},$
the horizon area becomes $A=8\pi M^{2}$ while $\widetilde{\Delta }=\left(
\widetilde{r}-M\right) ^{2}.$ The near horizon geometry is described by
setting
\begin{eqnarray}
\widetilde{r} &=&M+\lambda r \\
\widetilde{t} &=&\frac{t}{\lambda }  \nonumber \\
\widetilde{\phi } &=&\phi +\frac{t}{2M\lambda }  \nonumber
\end{eqnarray}

and taking the limit $\lambda \longrightarrow 0.$ One obtains (after
shifting also $r\longrightarrow \frac{1}{r}$ ) the geometry [7]
\begin{equation}
ds^{2}=\frac{\left( 1+\cos ^{2}\theta \right) }{2}\left[ \left( \frac{dt}{%
r_{0}r}\right) ^{2}-\left( \frac{r_{0}dr}{r}\right) ^{2}-\left( r_{0}d\theta
\right) ^{2}\right] -\frac{2\left( r_{0}\sin \theta \right) ^{2}}{\left(
1+\cos ^{2}\theta \right) }\left( d\phi -\frac{dt}{rr_{0}^{2}}\right) ^{2}
\end{equation}

in which $r_{0}^{2}=2M^{2}$ . We set further, for simplicity $%
r_{0}^{2}=2M^{2}=1.$ Although BR metric represents the throat region of an
extreme Reissner-Nordstrom BH the metric (4) \ due to itscross term signals
more of a BH property. It can easily be checked that the Killing vector $%
\partial _{t}$ becomes spacelike for $\cos 2\theta <4\sqrt{3}-7.$ We show
now that this near horizon geometry of an extreme Kerr BH coincides with the
colliding wave metric of CX [1]. The CX metric describing collision of
impulsive gravitational waves accompanied by shock waves is given by
\begin{equation}
ds^{2}=X\left( \frac{d\eta ^{2}}{\Delta }-\frac{d\mu ^{2}}{\delta }\right)
-\Delta \delta \frac{X}{Y}dy^{2}-\frac{Y}{X}\left( dx-q_{2}dy\right) ^{2}
\end{equation}

where
\begin{eqnarray}
X &=&\left( 1-p\eta \right) ^{2}+q^{2}\mu ^{2} \\
Y &=&p^{2}\Delta +q^{2}\delta  \nonumber \\
q_{2} &=&\frac{2q\left( 1-\eta \right) \left( p\mu ^{2}+p\eta +\mu
^{2}-1\right) }{\left( 1+p\right) \left( 1-p^{2}\eta ^{2}-q^{2}\mu
^{2}\right) }  \nonumber \\
\Delta &=&1-\eta ^{2}  \nonumber \\
\delta &=&1-\mu ^{2}  \nonumber
\end{eqnarray}

in which the constants of second polarization $(p,q)$ are constrained by
\begin{equation}
p^{2}+q^{2}=1
\end{equation}

We note that the metric function $q_{2}$\ is determined up to an additive
constant. This metric admits a horizon at $\eta =1$ and is free of curvature
singularities. Complete analytic extension of this space time was given in
detail by CX [1]. In particular the extended spacetime has timelike
singularities on hyperbolic arcs reminiscent of the ring singularity in the
Kerr space. Now let us choose $p=0,q=1$ so that the metric (5) becomes
\begin{equation}
ds^{2}=X\left( \frac{d\eta ^{2}}{\Delta }-\frac{d\mu ^{2}}{\delta }-\Delta
dy^{2}\right) -\frac{4Y}{X}\left( d\widetilde{x}-\eta dy\right) ^{2}
\end{equation}

where
\begin{eqnarray}
X &=&1+\mu ^{2} \\
Y &=&1-\mu ^{2}  \nonumber
\end{eqnarray}

and we have introduced the new coordinate
\[
d\widetilde{x}=\frac{1}{2}dx+dy
\]

Next we apply the transformation
\begin{eqnarray}
\eta &=&\frac{1}{2r}\left( r^{2}-t^{2}+1\right) \\
\tanh y &=&\frac{1}{2t}\left( r^{2}-t^{2}-1\right)  \nonumber \\
\mu &=&\cos \theta  \nonumber \\
\widetilde{x} &=&\phi -\frac{1}{2}\ln \left[ \frac{\left( r+t\right) ^{2}-1}{%
\left( r-t\right) ^{2}-1}\right]  \nonumber
\end{eqnarray}

which results in
\begin{equation}
ds^{2}=\left( 1+\cos ^{2}\theta \right) \left( \frac{dt^{2}-dr^{2}}{r^{2}}%
-d\theta ^{2}\right) -\frac{4\sin ^{2}\theta }{1+\cos ^{2}\theta }\left(
d\phi -\frac{1}{r}dt\right) ^{2}
\end{equation}

This metric is (up to a scale factor of $%
{\frac12}%
$ ) the line element (4) for the geometry of near horizon limit of an
extreme Kerr metric. We note that by scaling an overall constant $Q^{2}$ can
be introduced in front of this metric. This amounts, in the transformation
(10) to taking $\widetilde{x}\rightarrow Q\widetilde{x}$ and $y\rightarrow
Qy $. We add also that the choice $p=0,q=1$ pushes the timelike
singularities on hyperbolic arcs to infinity [1].

The Kerr-Newman metric has the same form as (1) except that \ $\widetilde{%
\Delta }=\widetilde{r}^{2}-2M\widetilde{r}+a^{2}+q^{2}$, where $q\ $ is the
electric charge and the $2M\widetilde{r}$ \ factor in $\omega $ is replaced
by $\widetilde{r}^{2}+a^{2}-\widetilde{\Delta }$ . The extreme Kerr-Newman
case corresponds to $M^{2}=a^{2}+q^{2}$ and the horizon is at $\widetilde{r}%
=M$ \ with area $4\pi \left( M^{2}+a^{2}\right) $ . Expanding \ $\omega $ to
first order in $\widetilde{r}-M$ \ one obtains
\begin{eqnarray}
\omega &=&\frac{a}{r_{0}^{2}}-\frac{2Ma}{r_{0}^{4}}\left( \widetilde{r}%
-M\right) \\
r_{0}^{2} &=&M^{2}+a^{2}  \nonumber
\end{eqnarray}

the near horizon limit takes the form [7]$\left( \text{with }%
r\longrightarrow \frac{1}{r}\right) $
\begin{eqnarray}
ds^{2} &=&\left( 1-\frac{a^{2}\sin ^{2}\theta }{r_{0}^{2}}\right) \left[
\left( \frac{dt}{r_{0}r}\right) ^{2}-\left( \frac{r_{0}dr}{r}\right)
^{2}-\left( r_{0}d\theta \right) ^{2}\right] \\
&&-r_{0}^{2}\sin ^{2}\theta \left( 1-\frac{a^{2}\sin ^{2}\theta }{r_{0}^{2}}%
\right) ^{-1}\left( d\phi +\frac{2Ma}{rr_{0}^{4}}dt\right) ^{2}  \nonumber
\end{eqnarray}

We proceed now to show that this corresponds to a metric of colliding cross
polarized em shock waves [8,9]. In the notation of CX this metric is given
by
\begin{equation}
ds^{2}=F\left( \frac{d\eta ^{2}}{\Delta }-\frac{d\mu ^{2}}{\delta }\right)
-\Delta \delta dy^{2}-\frac{\delta }{F}\left( dx-q_{0}\eta dy\right) ^{2}
\end{equation}

where
\begin{eqnarray}
2F &=&1+p_{0}+\left( p_{0}-1\right) \mu ^{2} \\
p_{0} &=&\sqrt{1+q_{0}^{2}}=const.  \nonumber
\end{eqnarray}

In terms of the null coordinates $(u,v)$ we have $\eta =sin(au+bv)$ and $\mu
=sin(au-bv)$, where $(a,b)$ are the constant em parameters. For $q_{0}=0$ $%
(p_{0}=1)$ this solution reduces to the BS solution of colliding em waves
with linear polarization. Applying a similar transformation as (10) (with $%
\widetilde{x}=x$ now) casts our metric into
\begin{equation}
ds^{2}=F(\theta )\left( \frac{dt^{2}-dr^{2}}{r^{2}}-d\theta ^{2}\right) -%
\frac{\sin ^{2}\theta }{F(\theta )}\left( d\phi -\frac{dt}{r}\right) ^{2}
\end{equation}

where
\begin{equation}
2F\left( \theta \right) =1+p_{0}+\left( p_{0}-1\right) \cos ^{2}\theta
\end{equation}

Comparing the metrics (13) and (16) we observe that upon choosing $q_{0}=-Ma$
with $M^{2}=a^{2}+2$ (which corresponds to a charge of $\sqrt{2}$ ) and
scalings $\phi \longrightarrow r_{0}^{2}\phi $, $t\longrightarrow \frac{t}{%
r_{0}^{2}}$ they become identical. This proves that the near horizon
geometry of an extreme Kerr-Newman BH is obtained as a result of two
colliding em waves with cross polarization.

A general proof, namely, that the extreme geometry of any BH is represented
by a CPW remains open. Generic representations of both classes ( of BHs and
CPWs ) may provide a general proof. We are satisfied here only with the two
simplest but most important cases. A general proof on the other hand is
reminiscent of a theorem by Penrose [10]: Any spacetime admits a plane wave
limit. Here it is not any spacetime but it is only the near horizon geometry
of an extreme BH. And it is not a plane wave limit but it is a colliding
wave limit

\begin{acknowledgement}
I wish to thank Ozay Gurtug for helpful discussions.
\end{acknowledgement}

\end{document}